\documentclass[manuscript]{aastex61}

\newcommand\aastex{AAS\TeX}

\accepted{to ApJ}

\shorttitle{\aastex\ L dwarf scale height}
\shortauthors{Sorahana et al.}
\begin{document}

\title{Evaluation of the Vertical Scale Height of L dwarfs in the Galactic Thin Disk
%\footnote{This version fixes many bugs from v6.0 and introduces some new features, primarily in the way the author and affiliations are now marked up.}
}

\author{Satoko Sorahana}
\email{satoko.sorahana@nao.ac.jp}
\affiliation{National Astronomical Observatory of Japan,
Mitaka, Tokyo 181-8588, Japan}

\author{Tadashi Nakajima}
\affiliation{Astrobiology Center,
Mitaka, Tokyo 181-8588, Japan}
\affiliation{National Astronomical Observatory of Japan,
Mitaka, Tokyo 181-8588, Japan}

\author{Yoshiki Matsuoka}
\affiliation{Research Center for Space and Cosmic Evolution,
Ehime University \\
Matsuyama, Ehime, 790-8577, Japan
}
%2000 Florida Ave., NW, Suite 300 \\
%Washington, DC 20009-1231, USA}
%\collaboration{(AAS Journals Data Scientists collaboration)}

%% Note that the \and command from previous versions of AASTeX is now
%% depreciated in this version as it is no longer necessary. AASTeX 
%% automatically takes care of all commas and "and"s between authors names.

%% AASTeX 6.1 has the new \collaboration and \nocollaboration commands to
%% provide the collaboration status of a group of authors. These commands 
%% can be used either before or after the list of corresponding authors. The
%% argument for \collaboration is the collaboration identifier. Authors are
%% encouraged to surround collaboration identifiers with ()s. The 
%% \nocollaboration command takes no argument and exists to indicate that
%% the nearby authors are not part of surrounding collaborations.

%% Mark off the abstract in the ``abstract'' environment. 
\begin{abstract}
 
Using the data release 1 of the Hyper Suprime-Cam
Subaru Strategic Program covering about 130 square degrees at high
galactic latitudes, we have obtained L dwarf counts based on the selection criteria
on colors, limiting magnitude and PSF morphology using $i$, $z$, and $y$ bands. 
3665 L dwarfs brighter than $z=24$ have been detected by these criteria. The surface number
counts obtained differentially in $z$ magnitude are compared with
predictions of an exponential disk model to estimate the thin-disk scale height
in the vicinity of the Sun.
In the exponential disk model, we first fix the local luminosity
function (LLF) to the mean LLF of Cruz et al. (2007) and 
derive the best fit scale height of 260 pc.
However this fit appears to be poor.
We then allow the LLF to vary along with the scale height.
We use the LLF of Cruz et al. as a starting point. 
%We use the number densities and their standard deviations
%of  seven magnitude bins of the LLF of Cruz et al. as 
%a starting point of searching for the optimum exponential disk model
%using a Monte Carlo technique. 
The best-fit model is found for the vertical scale height of 
380 pc.  However the $\chi^2$ minimum is rather broad and the 90\% confidence 
interval is between 320 and 520 pc.
We investigate another model by varying the scale height and the density of the brightest magnitude bin,
while other magnitude bins are fixed to the mean LLF of Cruz et al..
We  find  an equally good fit with the two free parameters  and the best-fit scale height 
is again 380 pc,
but the 90\% confidence interval is between 340 and 420 pc. 
\end{abstract}

%% Keywords should appear after the \end{abstract} command. 
%% See the online documentation for the full list of available subject
%% keywords and the rules for their use.

%% From the front matter, we move on to the body of the paper.
%% Sections are demarcated by \section and \subsection, respectively.
%% Observe the use of the LaTeX \label
%% command after the \subsection to give a symbolic KEY to the
%% subsection for cross-referencing in a \ref command.
%% You can use LaTeX's \ref and \label commands to keep track of
%% cross-references to sections, equations, tables, and figures.
%% That way, if you change the order of any elements, LaTeX will
%% automatically renumber them.

%% We recommend that authors also use the natbib \citep
%% and \citet commands to identify citations.  The citations are
%% tied to the reference list via symbolic KEYs. The KEY corresponds
%% to the KEY in the \bibitem in the reference list below. 

\section{Introduction} \label{sec:intro}

% scale height of low mass stars
% Chen 2001, Juric 2008, Bochanski 2010 (SDSS)
% Pirzkal 2009
% van Vledder 2016  Holwerda (Galactic coordinates are in error.) (HST)
% 
% Nakajima version of introduction
%

%It has been nearly 30 years, since the first L dwarf was discovered \citep{1988Natur.336..656B}
%and more than 20 years, since the first T dwarf was discovered \citep{1995Natur.378..463N}.
%These initial studies are of brown dwarf companions to stars, and their number
%is rather small. 
%It is now considered that these brown dwarf companions to stars are only a minority
%of the entire brown dwarf population.

The statistical study on the brown dwarf population has become possible by 
the advent of 
the large-area digital surveys such as  the Two Micron All Sky Survey (2MASS: 
\cite{2006AJ....131.1163S}),
Sloan Digital Sky Survey (SDSS: \cite{2000AJ....120.1579Y}), 
Canada-France brown dwarf survey and
its infrared version,
(CFBDS \& CFBDSIR: \cite{2010A&A...518A..39D}), UKIRT Infrared Deep Sky Survey (UKIDSS: \cite{2007MNRAS.379.1599L}) and a space-based all-sky survey by the Wide-field 
Infrared Survey Explorer (WISE: \cite{2010AJ....140.1868W}).
%By now, more than 1300 brown dwarfs are known  ({\bf Need reference}). These objects
%have provided us with a wealth of information about physics and chemistry
%in the photosphere of brown dwarfs \citep{1997ApJ...480L..39J,1999AJ....118.2466M,
%1999ApJ...519..802K,2000AJ....120..447K,2000AJ....119..369R,2000ApJ...536L..35L}. 
Utilizing the data from these large-area surveys,
several groups have determined  the number densities of various types of
brown dwarfs in the solar neighborhood, or more precisely the local luminosity
function (LLF) of the brown dwarfs
 \citep{2007AJ....133..439C, 2008ApJ...676.1281M, 2010A&A...522A.112R, 
2012ApJ...753..156K, 2013MNRAS.430.1171D, 2013MNRAS.433..457B, 2015MNRAS.449.3651M}.
%Derived LLFs coincide with each other within a factor of two to three (Figure 25 of 
%\cite{2015MNRAS.449.3651M}).

Beyond the immediate solar neighborhood, brown dwarfs are expected to have
density distribution similar to the stars in our Galaxy.
In other words, brown dwarfs will have thin-disk, thick-disk and halo populations, as stars do.
However, the determination of the spatial distribution of brown dwarfs in the context of
the Galactic structure is not an easy task because of their faintness.
In fact even the spatial distribution of
low mass stars, which are a little more massive than brown dwarfs, has been
determined  only recently.
The vertical scale height of low mass stars has been estimated 
by the analyses of the SDSS data  by several studies 
\citep{2001ApJ...553..184C,2008ApJ...673..864J,2010AJ....139.2679B}.
The derived scale height ranges from 290 to 330 pc.
The M dwarf scale height has been obtained also by the analyses of the HST data 
\citep{2009ApJ...695.1591P,2016MNRAS.458..425V} and it ranges
from 290 to 370 pc.
These results  from SDSS and HST should be compared with the spatial 
distribution of brown dwarfs.

% scale height of brown dwarfs
% Pirzkal 2005, Ryan 2005, Ryan 2011
The previous ground-based surveys are not deep enough to detect 
brown dwarfs as far as 300 pc, which is roughly 
the scale height of low mass stars. Only data that reach beyond 300 pc,
are obtained by the HST, whose pencil beam surveys have picked up 
rather small numbers  of brown dwarfs, but have given estimates of
the vertical scale height, which range from 290 to 400 pc
\citep{2005ApJ...622..319P,2005ApJ...631L.159R,2011ApJ...739...83R}.

The Hyper Suprime-Cam Subaru Strategic Program (HSC-SSP) survey is capable
of detecting a large number of 
new  brown dwarfs because of a greater limiting distance and a larger
volume for a given spectral type compared with previous surveys. 
%In \citet{2016ApJ...828...26M,2018PASJ...70S..35M,2018ApJS..237....5M}, 
%they found fifteen new L and T dwarfs from
%HSC-SSP data and spectroscopic follow up using the Gran Telescopio Canarias and
%Subaru.
In this paper, we report the number count of new color-selected L dwarfs
using the HSC-SSP survey
data and discuss the vertical distribution of L dwarfs in the Galactic thin disk.

Our paper is organized as follows. In section 2, we provide the information on
HSC-SSP data and L dwarf selection in our analysis. 
In section 3, we show the results of comparison of the magnitude dependent
number counts of L dwarfs obtained from the HSC-SSP data and predictions
of an exponential disk model. 
In section 4, our results are compared with the previous works on 
brown dwarfs and low mass stars.

\section{Data and L Dwarf Selection} \label{sec:data}

\subsection{HSC DR1 Data} \label{subsec:dr1}

HSC is a new prime-focus camera mounted on
the 8.2 m Subaru telescope \citep{2018PASJ...70S...1M}. Compared with the
Suprime Cam, the prime-focus imager  of the previous generation, the
field of view (FOV) of the HSC is ten times larger, while it is less affected by
distortion. HSC has 104 red-sensitive science CCDs whose 870 Mega pixels
cover 1.5 degree diameter FOV with a 0.168 arcsec pixel scale. 
HSC-SSP,
which is led by the astronomical communities of Japan, Taiwan,  and
Princeton University,
is carrying out a multi-band ($g,r,i,z$, and $y$ broad bands, and 
4 narrow-band filters) imaging survey.
The survey consists of three layers with different depths: wide, deep and ultra-deep.
We use 130 square degrees of the data release 1 (DR1) of the wide  and deep surveys,
which was released in February 2017 \citep{2018PASJ...70S...8A}. 

%The wide survey will
%eventually cover $\sim$1400 deg$^2$ in $g, r, i, z$ and $y$ bands to the depth
%of $i\sim25.8$ with 5 $\sigma$ significance.

Several astronomical surveys related to brown dwarf research have been carried out 
so far. The representative ground-based surveys are 
2MASS  \citep{2006AJ....131.1163S},
SDSS,
\citep{2000AJ....120.1579Y}, CFBDS \& CFBDSIR, 
\citep{2010A&A...518A..39D}, UKIDSS \citep{2007MNRAS.379.1599L},
Visible   and Infrared Survey Telescope for Astronomy (VISTA) 
Kilo-Degree Infrared Galaxy Survey and
Hemisphere Survey
(VIKING \& VHS: \cite{2007Msngr.127...28A}). There is also 
a space-based all-sky survey, WISE \citep{2010AJ....140.1868W}.

\begin{deluxetable}{lcccccccr}
%\rotate
\tabletypesize{\scriptsize}
  \tablecaption{Comparison of limiting distances and survey volumes \label{tab:table1}}
\tablewidth{0pt}
\tablehead{
\colhead{survey}&\colhead{SpT}&\colhead{$M_z$}  &\colhead{limiting $z$ mag}&\colhead{$M_J$}&\colhead{limiting $J$ mag}&\colhead{limiting $d$}&\colhead{survey area}&\colhead{volume}\\
&&&&&&\colhead{[pc]}&\colhead{[deg$^{2}$]}& \colhead{[pc$^{3}$] }  \\}
\startdata
HSC&L5&16.4&24.3&--&--&378&1400& $7.70\times10^6$ \\
HSC&T6&17.8&24.3&--&--&196&1400& $1.07\times10^6$ \\
SDSS&L5&16.1&20.0&--&--&60&10000& $2.19\times10^5$ \\
SDSS&T6&17.5&20.0&--&--&31&10000& $3.04\times10^4$ \\
CFBDS&L5&16.1&22.5&--&--&190&600&  $4.16\times10^5$\\
CFBDS&T6&17.5&22.5&--&--&98 &600 & $5.76\times10^4$\\
CFBDSIR & L5 & -- & -- & 13.8 & 20.0 & 177 & 335 & $1.89\times10^5$\\
CFBDSIR & T6 & -- & -- & 15.1 & 20.0 & 97 & 335 & $3.09\times10^4$ \\
UKIDSS&L5&--&--&13.8&19.3&128 &4000& $8.56\times10^5$\\
UKIDSS&T6&--&--&15.1&19.3&70 &4000& $1.40\times10^5$ \\
VHS      &L5&--&--&13.8&19.4&  134     &20000& $4.92\times10^6$  \\
VHS      &T6&--&--&15.1&19.4&   73     & 20000 &   $8.05\times10^5$ \\
VIKING     &L5&--&--&13.8&20.4&  213     & 1500 & $1.47\times10^6$  \\
VIKING     &T6&--&--&15.1&20.4&   116    & 1500 &   $2.40\times10^5$ \\
2MASS&L5&--&--&13.8&15.8&26&40000& $6.80\times10^4$ \\
2MASS&T6&--&--&15.1&15.8&14 &40000& $1.11\times10^4$\\
\enddata
\tablecomments{Limiting magnitudes are given for 10-sigma significance.
$z$ magnitude is in the AB system, while $J$ magnitude is in the Vega system.
$M_z$ and $M_J$ are from \citet{2015ApJ...810..158F}.
$M_z(hsc) - M_z(SDSS) = 0.3$ (This paper). }
\end{deluxetable}

\begin{figure}
%\vskip -4cm
%\epsscale{.45} 
\begin{center}
%\plotone{comparison.eps}
\includegraphics[width=12cm,angle=0]{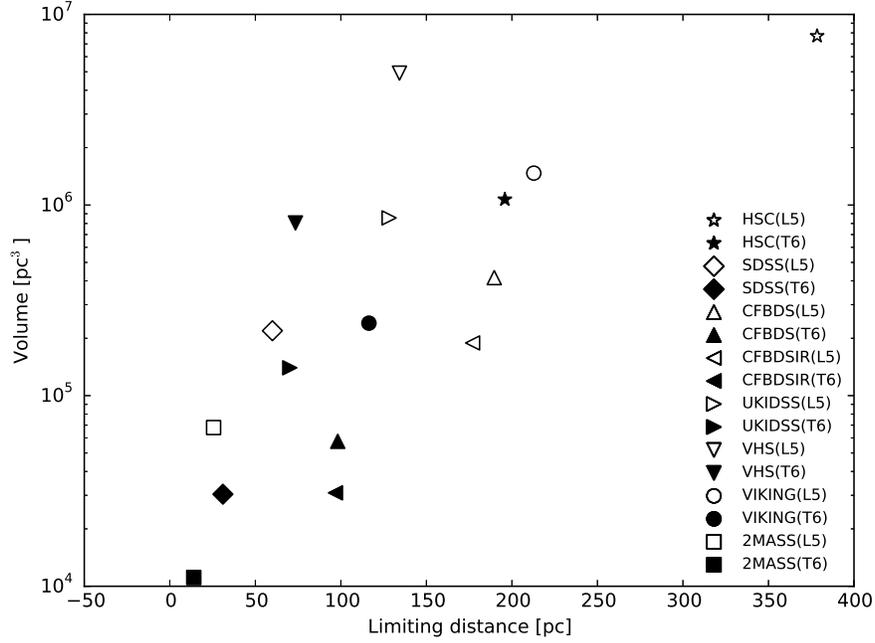}
\end{center}
%\vskip 3cm
\caption{The limiting distances and volumes for L5 and T6 dwarfs for different surveys. 
For L5 dwarfs,
the HSC survey is the only survey that reaches beyond 300 pc, which is about the scale height
of M dwarfs.
} 
\label{fig:comparison} 
\end{figure}

We summarize the 10-sigma limiting magnitudes of $z$ band for HSC, SDSS, and
CFBDS, and of $J$ band for 2MASS, UKIDSS, CFBDSIR, VHS and 
VIKING, limiting distances in units
of pc, survey areas in units of deg$^2$, and volumes in units of pc$^3$ for L5
and T6 dwarfs in Table \ref{tab:table1}. 
WISE is sensitive at 4.6 $ \mu$m and main targets are late T and Y dwarfs
within 10 pc of the Sun \citep{2011ApJS..197...19K} and therefore we have excluded WISE
from this comparison.
Figure \ref{fig:comparison} shows the limiting distances and survey volumes
for L5 and T6 dwarfs for different surveys.  For L5 dwarfs,
HSC is the only survey that reaches more than 300 pc, which is approximately
the scale height of M dwarfs. 
VIKING is the runner-up, while the volume covered by
VHS will be comparable to  that covered by HSC at the time of their completion.
For T6 dwarfs,  HSC reaches almost 200 pc and 
VIKING is again in the second place  ($\sim$ 120 pc),
though neither of the surveys reaches 300 pc, which is about the necessary 
limiting distance to obtain a scale height.

\subsection{Deep and Wide  fields}

We use three fields of the Deep layer and six fields of the Wide layer.
The name and properties of each field are given in Table \ref{tab:obs}.
The major difference between the Deep and Wide fields is that
narrow-band filters were used in addition to broad-band filters in
the Deep layer.  As far as broad-band imaging is concerned,
the difference in depth in  each band is 0.1 mag \citep{2018PASJ...70S...8A},
which is much smaller than  0.42 mag, which is the uncertainty
in the absolute $z$ mag of L dwarfs, which appears in model fitting procedure (Section 3).
The survey areas of the Wide fields in general are greater than those of
the Deep fields, but there is one exception, HECTOMAP (Wide), whose
survey area is smaller than those of the three Deep fields.
For these reasons, we treat all nine fields in the same manner
in the following analysis.

It should be emphasized that the above statements regarding the depth of
the Deep survey and the width of HECTOMAP are strictly true for
the DR1 release, since the surveys are still on-going.
 At the time of completion, the Deep survey will be much deeper
and the area of HECTOMAP will be much wider.

\subsection{Selection conditions}
We first choose data sources from the HSC-SSP DR1 database for non-blended sources\footnote{The sources isolated or deblended from parent blended sources. We reject those parents. This corresponds to ``$deblend.nchild = 0$" in the database language.}  satisfying the following conditions:

\begin{enumerate}
\item{$S/N > 10 $ in $z$}
\item{$1.0 <i-z < 2.0$}
\item{$0.75 < z-y < 1.0$}
\item{ $z -  z_{cModel} < 0.15$}
\end{enumerate}

where $i, z$, and $y$ are in the AB magnitude system. Here magnitudes of the HSC
data refer to point-spread function (PSF) magnitudes. 
$z_{cModel}$ is measured by fitting the PSF-convolved galaxy models to the source profile. 
The difference between the PSF and $cModel$ magnitudes is used to exclude extended sources.

In addition, we use the following critical quality flags, which are used in \citet{2016ApJ...828...26M, 2018PASJ...70S..35M,2018ApJS..237....5M}: we require that the source (i) is not close to an edge of the image frame (flags$\_$pixel$\_$edge = False), 
(ii) is not in a bad CCD region (flags$\_$pixel$\_$bad = False), 
(iii) is not saturated (flags$\_$pixel$\_$saturated$\_$center = False),  (iv) is not affected by cosmic rays (flags$\_$pixel$\_$cr$\_$center = False)  in the $i$, $z$, and $y$ bands,  
(v) does not include a footprint in bright object pixels 
(zflags$\_$pixel$\_$bright$\_$object$\_$center =   False).
Below we discuss the implications of each selection criterion.

\subsection{L dwarf selection}

\noindent
Condition 1.  

$z$ band is the primary band where the observed and model counts are compared 
in section 3. This condition for the S/N is effectively equivalent to $z<24$ and $\sigma_z < 0.1$ mag.
The photometric uncertainty of $\sigma_z < 0.1$ is much smaller than the intrinsic scatter
of the absolute $z$ magnitude of the L dwarfs of 0.42 mag, which we derive in section 3.
Interstellar extinction of the observed fields ranges from $A_z = 0.03$ to 0.09
\citep{2011ApJ...737..103S}, which is again much smaller than the intrinsic scatter of the absolute
$z$ magnitude of L dwarfs. So we neglect the effect of interstellar extinction from our analysis.
We consider that the observed L dwarf count is complete to $z=24$ as we show in 
section  \ref{completeness}.

\noindent
Conditions 2 \& 3.
 
\begin{figure}
%\vskip -4cm
%\epsscale{.45} 
\begin{center}
\plotone{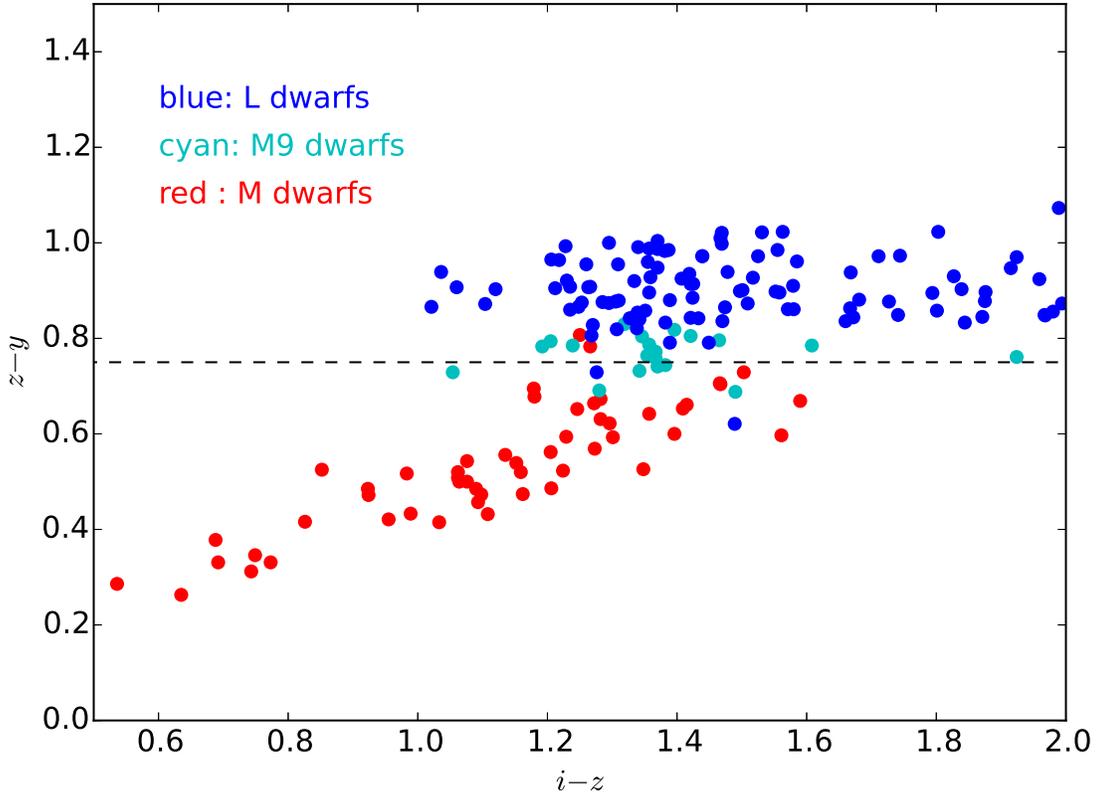}
\end{center}
%\vskip 4cm
\caption{The $i-z$ vs. $z-y$ diagram. The known M($\le 8$),  M9(including M8.5), and L dwarfs are shown in red, cyan, and blue points, respectively.  The M8/M9 border ($z-y$ = 0.75) is shown in black dashed line. 
This border is selected so that early L dwarfs are included even if 
some of the M9 dwarfs are also included.
}
\label{fig:color-color}
\end{figure}

Figure~\ref{fig:color-color} shows the $i-z$ vs. $z-y$ diagram for known
M dwarfs (red), M9 dwarfs (cyan), and L dwarfs (blue).
The $i-z$ and $z-y$ colors of these dwarfs are synthesized from the HSC filter functions of $i$, $z$, and $y$ bands for spectra obtained with SpeX, a medium-resolution spectrograph covering 0.65--5.4~$\mu$m at the NASA Infrared Telescope Facility (IRTF)\footnote{We retrieve the data from the SpeX Prism Spectral Libraries built by Adam Burgasser. http://pono.ucsd.edu/$\sim$adam/browndwarfs/spexprism/}, and CGS4, a cassegrain spectrograph at the United Kingdom Infrared Telescope (UKIRT)\footnote{We retrieve the data from the L and T dwarf data archive built by Sandy Leggett. http://staff.gemini.edu/$\sim$sleggett/LTdata.html}. 
These colors of known brown dwarfs are derived by 
\citet{2016ApJ...828...26M, 2018PASJ...70S..35M,2018ApJS..237....5M}.

$i-z=1.0$ is the blue border of L dwarfs, while $i-z=2.0$ is almost the red border
of L dwarfs. 
It should be noted that the $z(hsc)$ filter transmits rather shorter wavelengths than the 
$z^\prime(SDSS)$ 
filter, since the $y$-band is used as the longest wavelength band. 
In Figure 1 of \citet{2016ApJ...828...26M}, an $i-z$ vs $z-y$ digram is plotted
for wider ranges of $i-z$ and $z-y$. According to this figure, L/T transition occurs 
around $i-z \sim 2.5$ and we exclude the latest L dwarfs (roughly L9) and 
L/T transition objects by the criterion $i-z<2.0$.
The reason for the exclusion of L9 and later dwarfs is that our L dwarf count
should be consistent with the local luminosity function (LLF) 
we use of \citet{2007AJ....133..439C}, which
includes L dwarfs down to L8, but does not include L9 and L/T transition objects.

Compared with the L/T border, the M/L border is more problematic. 
We set $z-y = 0.75$ as an M/L border around which M/L transition objects are mixed up.
We can see that almost all of the M-type objects included in the M/L mixed zone around
$z-y \sim 0.75$ are M9(and M8.5)  dwarfs. Since it is intrinsically difficult to determine the spectral class
for the M/L borderline objects, we regard $z-y=0.75$ as an M8/M9 border. So we allow
M9 dwarfs to be included in the L dwarf sample, while we avoid excluding early L dwarfs from
the L dwarf sample. M9 dwarfs are taken into account in the brightest magnitude bin
of the LLF of Cruz et al. in section 3.

\subsection{Completeness \label{completeness}}

\begin{figure}
%\vskip -4cm
%\epsscale{.45} 
\begin{center}
\plotone{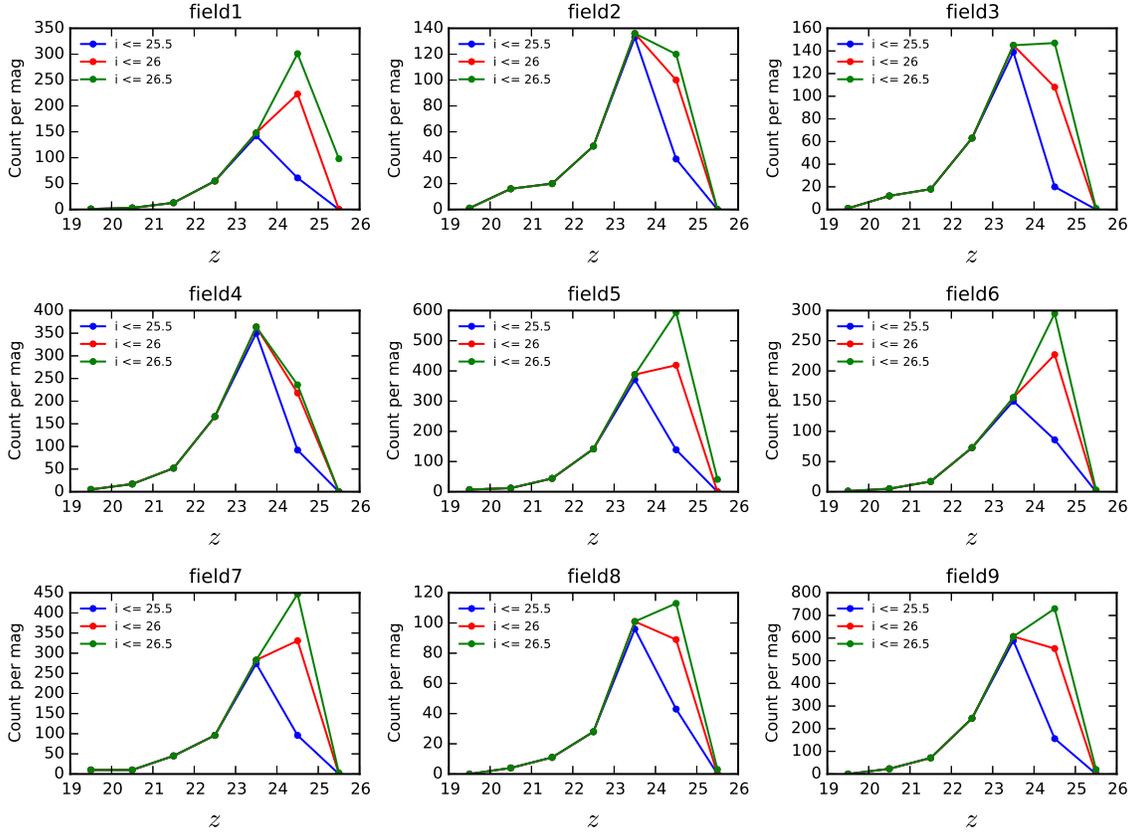}
\end{center}
%\vskip 4cm
\caption{$z$ band source counts are compared for $i<25.5, 26.0$ and $26.5$ for
each of the nine fields. The three source counts for each field do not differ one another down to
$z = 23.5$ or (23-24) bin.  These results combined with the color criterion $i-z < 2.0$ 
indicate that our sample is complete to $z=24$. The names of the fields are given in Table \ref{tab:obs}.
}
\label{fig:completeness}
\end{figure}

%\clearpage

Since $i-z < 2.0$, the faintest and reddest object at $i$ selected by the criterion $z<24$ is 
$i = 26$. In Figure \ref{fig:completeness}, differential source counts per $z$ magnitude
for all nine fields that satisfy the selection criteria 
are plotted for $i<25.5, 26.0$ and 26.5. The source counts of the three
cases agree reasonably well down to $z=23.5$ or (23-24) bin, while they deviate from $z=24.5$.
It is essential that the source counts agree down to $z=23.5$ for $i=26$ and $i=26.5$.  
From $z-y > 0.75$ and $z<24$,  $y < 23.25$ for which S/N $>8$,
and the sources at $y$ band have higher S/N than at $i$ band.
From the interpretation of Figure \ref{fig:completeness}, we confirm that
the source counts are complete to $z=24$ for $i-z < 2.0$.

Another potential source of incompleteness is the seeing dependence of
the point-source selection condition 4, $z -  z_{cModel} < 0.15$.  Seeing variation
exists among different fields and also within a field. 
In their high-redshift quasar survey, Matsuoka et al. (in preparation)
investigated the magnitude dependent completeness
due to the point-source selection, taking into account the seeing variation. According to their
formula,
the completeness was 99 and 95\% respectively at $z=22.5$ and 23.5, regardless of seeing.
In the model fitting procedure described in Section 3, we compared
the case for raw observed counts and that for completeness corrected counts.
It was found that the effect of correction was not negligible at the level of
20 pc difference in the scale height. So we included completeness correction 
due to point-source selection.

\subsection{Potential contaminants}

\subsubsection{L subdwarfs}

In this paper, we consider only thin disk population of L dwarfs
and we neglect thick disk and halo populations of L dwarfs, or so called L subdwarfs (sdL).
To justify our approach, here we estimate the contribution of  L subdwarfs in two ways.

$i -z$ and $z-y$ colors of L subdwarfs are not known, but we assume that
all L subdwarfs satisfy our color criteria and estimate an upper limit of L subdwarf contribution.
$r - z$ color of subdwarf M8 is known to be $\sim$ 2.8 \citep{2008ApJ...681L..33L} and
we assume that $r-z \sim 3$ for sdL0.
\citet{2017A&A...598A..92L} surveyed 3700 deg$^2$ of SDSS and UKIDSS data
down to $r\sim 23$ or approximately $z\sim20$, and detected 3 L subdwarfs (two sdL0 and one sdL0.5).
In terms of surface number density, this corresponds to 8.1$\times 10^{-4}$ per deg$^2$.
If we estimate the surface number density for the HSC limit of $z=24$,
it will be 0.21 per deg$^2$. If we apply the completeness correction as Lodieu et al did,
the surface number density increases to 0.51 per deg$^2$. For our survey area of some 130 deg$^2$,
we expect an upper limit of 66 L subdwarfs which is less than 2\% of our total L dwarf count.
We consider this contribution to be negligible.

Since \citet{2017A&A...598A..92L} detected only very early L subdwarfs, one may worry about
the contribution of L subdwarfs of later spectral type.  About the entire L subdwarf population,
some hint can be obtained from the result of the ALLWISE sky survey by \citet{2014ApJ...783..122K}.
According to Table 6 of their paper, the volume number density in
the immediate solar neighborhood of L subdwarfs  can be estimated to be about 0.3\% of that of thin disk L dwarfs.
If we move out of the Galactic plane by one exponential scale height, the fraction will increase to
0.9\%, but this fraction is still small. Again we conclude that L subdwarf population is negligible.

\subsubsection{Passive elliptical galaxies at z $\sim 1$}

Passively evolving elliptical galaxies at z(redshift)$\sim1$ have large $i - z$ colors
due to the 4000{\AA} break in the rest-frame SEDs. 
\citet{2005ApJ...634..861Y} detected about 4000 such objects (z=0.9 $\sim$ 1.1) in 1.03 deg$^2$
of the Subaru/XMM-Newton Deep Survey,  which are brighter than $z^\prime < 25$
using the Suprime Cam, the predecessor of HSC.
There may be more than 1000 passively evolving elliptical galaxies brighter than $z=24$
per deg$^2$, which satisfy $i-z$ color criterion. 
If selection criteria other than $i-z>1.0$ could not reject these galaxies, they would outnumber 
L dwarfs.

Here we argue that the color criterion $z - y > 0.75$ efficiently
excludes these galaxies based on the following reasoning.
At z$\sim 1$, the rest-frame wavelengths of $z$ and $y$ are
4400 and 4800{\AA} respectively. The rest-frame SEDs of passive elliptical galaxies
of various ages and metallicities are found in \citet{1989ApJ...337..125B}. 
The ratio of rest-frame flux densities
$f_\nu(4400)/f_\nu(4800)$ ranges from 1.07 to  0.83 from the  most metal poor to
the extremely metal rich case.
 Since the color in AB magnitude is
effectively the ratio of the flux densities, which is preserved even if
the redshift is changed,  $z - y$ ranges from -0.07 to 0.2.

Now we consider if the intrinsically reddest galaxy with $z-y=0.2$ accidentally
has $z-y>0.75$  due to photometric errors in $z$ and $y$.
The reddest and faintest galaxy will have the intrinsic $z = 23.25$ and $y=23.0$.
1 sigma photometric errors for this galaxy are 0.05 and 0.1 respectively for
$z$ and $y$ and corresponding 1 sigma error in $z-y$ is 0.11.
Since $z - y = 0.75 = 0.2 + 5 \times 0.11$,
the photometric error in $z-y$ needs to exceed 5 sigma level in order to
contaminate L dwarfs. It should be emphasized that intrinsic $z-y$ of
most of the galaxies is less than 0.2.
Passive elliptical galaxies at z$\sim1$ are not likely to be as red as
L dwarfs in $z - y$.

In addition to the $z-y$ color criterion, the PSF criterion (condition 4) 
discussed in Section \ref{completeness}, is
efficient in excluding passive elliptical galaxies which are expected to be extended.
This is another reason that
passive elliptical galaxies are not significant contaminants.

\subsubsection{High-redshift quasars and high-redshift galaxies}\label{highzqso}

Since \citet{2016ApJ...828...26M,2018PASJ...70S..35M} conducted a high redshift quasar survey
using the HSC-SSP Wide Survey data based on color-color selection
criteria similar to ours, their results are most relevant to our analysis.

In Figure 1 of \citet{2016ApJ...828...26M}, they plot in the $i-z$ vs $z-y$ diagram
the loci of high-redshift quasars and high-redshift galaxies as well as the locations
of L and T dwarfs. In this color-color diagram, quasars and galaxies overlap
with L dwarfs for only a narrow range of z(redshift) $\sim 6.5$. 
In this redshift interval, the expected number
of quasars with $z < 24$ is a few in 130 deg$^2$ and that of galaxies which
appear to be point like in HSC images is also a few in 130 deg$^2$
 \citep{2016ApJ...828...26M,2018PASJ...70S..35M,2018ApJS..237....5M}.
These results imply that both quasars and galaxies are negligible for our
L dwarf count.
 
We have finally extracted a total of 3665 sources in which only a very small fraction of
objects are expected to be non thin-disk L dwarfs.
%In Figure~\ref{fig:color-color}, HSC sources are plotted in gray points. 
%We find that most of objects are placed on the $i-z$ vs. $z-y$ sequence of brown dwarfs, especially on %the L dwarf sequence.

\section{L dwarf distribution in the Galactic disk} \label{sec:BDdistribution}

\subsection{Exponential disk model}

\subsubsection{Thin disk model}

Here we  consider only  the thin disk population by an exponential disk model.
The number density of the L dwarfs of
the $k$ th  magnitude bin of the $J$ band luminosity function
at the Galactic position $(R,Z)$
in cylindrical coordinates, is given by

\begin{equation}
\Phi_k(R,Z) = \Phi_k  \exp\left(\frac{R_0-R}{H} -\frac{|Z|}{h} \right),
\end{equation}

where $\Phi_k$ is the number density of
the $k$ th magnitude bin of the $J$ band local luminosity function (LLF)
 of the L dwarfs of the immediate solar neighborhood,
 $R_0 = 8$ kpc is the Galactocentric distance, $H = 3.5$ kpc is the 
Galactic scale length, and $h$ is the vertical scale height to be estimated
from the comparison of the model and observations. 
We start from $J$-band because a L dwarf LLF has been  obtained previously at $J$ band
by \citet{2007AJ....133..439C}.
The range of absolute $J$ magnitude is from 11.75 to 14.75 with a 0.5 mag step, covering
in spectral type from M9/L0 to L8.
In the calculation of the surface number density in the line of sight of the Galactic coordinates $(l,b)$,
the height of the Sun above the Galactic plane, $Z_\odot = 27$ pc \citep{2001ApJ...553..184C} is taken into account.

\subsubsection{Error in $z$-band absolute  magnitude}

The absolute $J$ magnitude
is converted to absolute $z(hsc)$ magnitude by the $z(SDSS)-J$ vs. $M_J$ relation
\citep{2015ApJ...810..158F} and $z(hsc) - z(SDSS) = 0.3$ which we derived.
The relation between $M_J$  and $z(hsc)-J$ is given in Table \ref{tab:zmJ}.  
The uncertainty involved in this conversion is estimated to be 0.3 mag.

In the work of Cruz et al., the uncertainty in spectral type is a half subtype,
which induces an uncertainty of 0.3 mag in $M_J$.
Assuming that the error in the color conversion and that in $M_J$
are independent, the total error in the absolute $z$ mag amounts to
0.42 mag, which
 causes significant Malmquist bias. Photometric error is less than 0.1 mag and
 this contribution is small compared with the error in $M_z$.
 In Appendix, our treatment of Malmquist bias following the original procedure
 adopted by \citet{1922MeLuF.100....1M} is given. The only difference in our case is
 that we used numerical integration in apparent $z$ magnitude which is usable
 for any form of the density distribution,
 while Malmquist used analytic integration for the cases it is possible (The famous formula for example
 is for a uniform density distribution).

\begin{table}
\begin{center}
%\caption{$M_J$ vs. $z(hsc)-J$ \label{tab:zmJ}}
\begin{tabular}{cccccccc}
\tableline\tableline
$M_J$ & 11.75   & 12.25  &  12.75 & 13.25 & 13.75 & 14.25 & 14.75  \\
$z(hsc)-J$  & 2.25  &  2.40  &   2.52 & 2.61 & 2.65 & 2.70 & 2.75 \\ 
\tableline
\end{tabular}
\end{center}
\caption{$M_J$ vs. $z(hsc)-J$ \label{tab:zmJ}. 
$M_J$ vs. $z(SDSS)-J$ relation was obtained from \citet{2015ApJ...810..158F} and
$z(hsc) - z(SDSS)=0.3$ was derived by us.
The uncertainty in this color-magnitude
 relation induces 0.3 mag uncertainty in $M_z$,
which is taken into account in the analysis. }
\end{table}

\subsection{Fitting observational data with a model} \label{subsec:model}

Since we have more than 3600 sources, we choose to use a differential number count
instead of the cumulative number count so that
the error estimate in each magnitude bin is more strict.
We have 45 degrees of freedom (5 magnitude bins $\times$ 9 fields)
in our observing data.

A star count model that gives the surface number of stars
per magnitude as a function of apparent magnitude
for Galactic coordinates of the field of concern
is a function (or functional) of LLF and the vertical scale height $h$.
We denote  the model number count as $M(LLF,h:m,f)$, where $m$ and $f$ indicate
the apparent magnitude bin and the field respectively. 
Here $f$ specifies
the area and the Galactic coordinates of each field.
The observed star count $O(m,f)$ is specified by the apparent magnitude bin and
the observed field as given in Table \ref{tab:obs}. The completeness correction
described in section \ref{completeness} is applied to magnitude bins $m$=22.5 ($22-23$)
and 23.5 ($23-24$).

\begin{deluxetable*}{ccccccccccc}[b!]
\tablecaption{L dwarf count differential in $z$ mag in each field \label{tab:obs}}
\tablecolumns{10}
%\tablenum{2}
\tablewidth{0pt}
\tablehead{
\colhead{Field} & \colhead{Layer}  & \colhead{Name}  & \colhead{l} & \colhead{b} & \colhead{area} &
\colhead{} & \colhead{} &  \colhead{$z(hsc)$} & \colhead{}   \\
 & & & \colhead{deg} & \colhead{deg} & \colhead{deg$^2$} & 19-20 &20-21 & 21-22 & 22-23 & 23-24 
}
\startdata
1 & DEEP & E-COSMOS & 237 & +43 &  8.6 & 1 & 3 & 13 & 55 & 148 \\
2 & DEEP & ELAIS-N1    &  85   & +44  &  9.2 &  1 & 16 & 20 & 49 & 136 \\
3 & DEEP & DEEP2-3    &  84   & -56   &  9.3 & 1  & 12 & 18 & 63 & 145 \\
4 & WIDE & XMM-LSS   & 170  & -59   &  23.9 & 5 & 17 & 52 & 166 & 364 \\
5 & WIDE & GAMA09H  & 228 & +28     &  18.6 & 7 & 12 & 44 & 142 & 388 \\
6 & WIDE & WIDE12H   & 278 & +60     &  12.3 & 1 &  5  & 17 &  73  &  156 \\
7 & WIDE & GAMA15H  & 349 & +53     &   16.5 & 10 & 10 & 45 & 96 & 283 \\
8 & WIDE & HECTOMAP & 69 & +45     &   6.2   & 0  & 4    &  11 & 28 & 101 \\
9 & WIDE & VVDS          &  66  & -45      & 25.1   & 0 & 23 & 71 & 246 & 607 \\
\enddata
\tablecomments{These differential number counts are raw counts.
In the $\chi^2$ fitting, completeness corrections of 1.3 and 5.4\% are applied
to magnitude bins, (22-23) and (23-24) respectively.
}
\end{deluxetable*}

The goodness of fit $\chi^2(LLF,h)$ is defined by

\begin{equation}
\chi^2(LLF,h) = \sum_f \sum_m \frac{|O(m,f) - M(LLF,h:m,f)|^2}{M(LLF,h:m,f)}.
\end{equation}

%As we have discussed in the previous section, the number count is
%complete to $z = 23-24$ bin. So five magnitude bins, $(19-20),(20-21), (21-22), (22-23)$
%and $(23-24)$ are used for the fitting procedure and the degrees of freedom (DOF) for data points
%are 30 (6$\times5$).

%It should be emphasized that we use the number count differential in magnitude,
%since we have sufficient information in each magnitude bin 
%due to the large number of sources so that we do not
%lose the details of the behavior of the number densities by using a cumulative
%number count. 

\subsection{Model for a fixed local luminosity function} \label{subsec:meanLF}

\begin{figure}
\begin{center}
%\plotone{cmp1m220.eps}
\includegraphics[width=12cm,angle=-90]{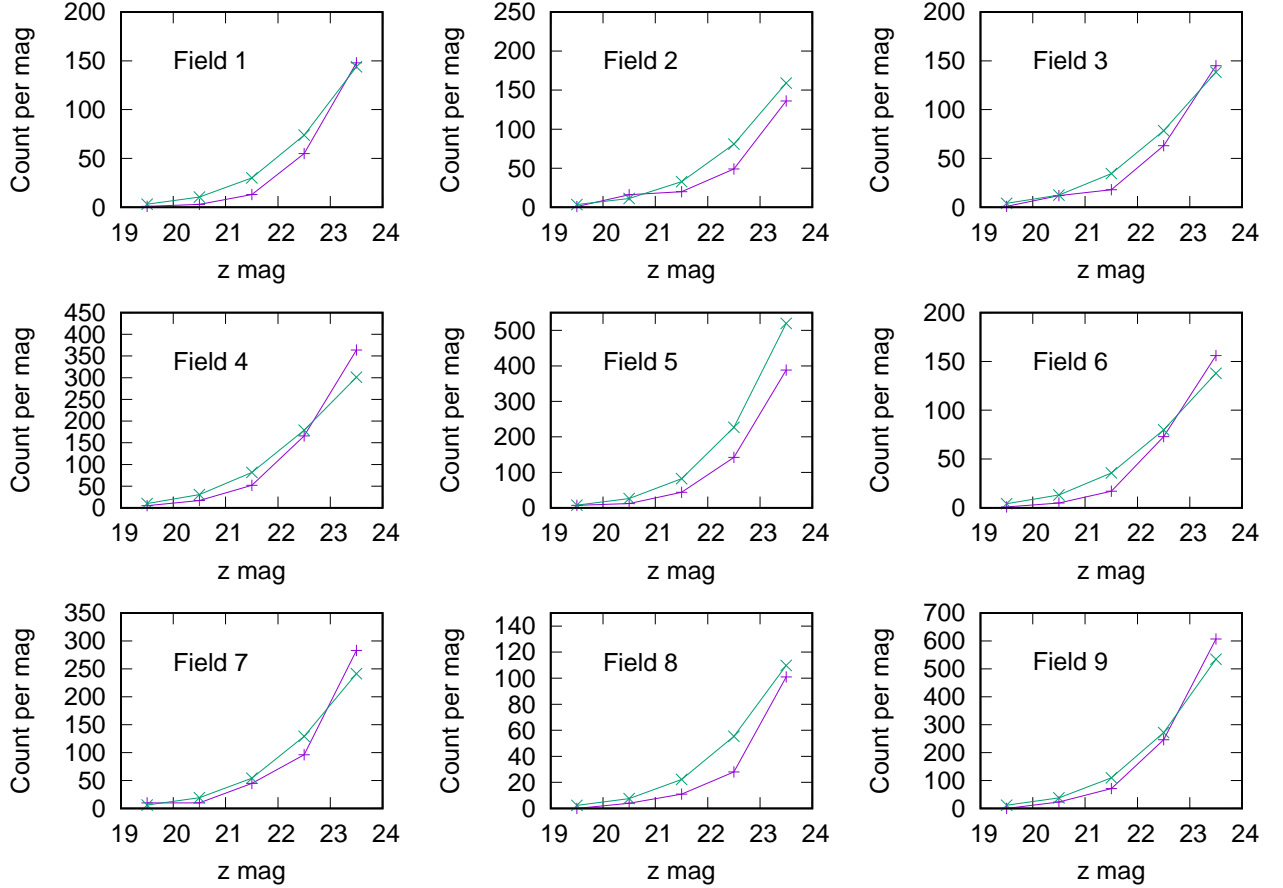}
\end{center}
\caption{Comparison of observed counts (violet) and model predictions (green) for
the mean LLF of Cruz et al.  (MCRUZ). 
The observed counts are corrected for completeness.  
The optimal $h= 260$ pc and $\chi^2 = 309$ for DOF = 44. The figure of merit
$Q = 1.4\times10^{-80}$ is very small and this fit is considered to be very poor.
} 
\label{meanLF}
\end{figure}

First we fix the LLF to the mean LLF of Cruz et al. and minimize the $\chi^2$
by varying $h$ from 100 to 480 pc with a 20 pc step.  
"mean LLF" implies that  the number densities of the individual magnitude bins are fixed
to the mean number densities and their standard deviations are ignored.
The minimum of $\chi^2$, $\chi^2_{min}$ = 309 
for 44 ($=45-1$) DOF is obtained for $h = 260$ pc.
The observed L dwarf counts  and model predictions are plotted in Figure \ref{meanLF}.

At a glance, the fit appears rather poor. However, we should be objective
in judging the goodness of fit. From now on, we compare the goodness of fit for
different degrees of freedom. The proper procedure for comparison is to use
$\chi^2$ distribution for $N$ degrees of freedom $f_N(\chi^2)$ and calculate
the cumulative probability, 
$P(N|\chi^2) = \int_0^{\chi^2}f_N(\chi^2) d\chi^2$. In our case, $P$ is close to unity
and a more convenient figure of merit is $Q(N|\chi^2) = 1 - P(N|\chi^2)$.
Larger the $Q$, better the fit.
For the present one-parameter fitting, $Q(44|309) = 1.4 \times 10^{-80}$,
which should be compared with the cases when we vary the LLF in the 
LLF in the following. From now on, we call the mean LLF as MCRUZ
and the best-fit model for the MCRUZ LLF as the MCRUZ model.

\begin{deluxetable*}{ccccc}[b!]
\tablecaption{$J$-band local luminosity functions  \label{tab:LF}}
\tablecolumns{5}
%\tablenum{2}
\tablewidth{0pt}
\tablehead{
\colhead{} & \colhead{CRUZ7} & \colhead{ITERA} & \colhead{FINAL}  &\colhead{HYBRD}\\
\colhead{$M_J$ } &  \multicolumn{4}{c}{$\Phi_k$} \\
\colhead{}             &  \multicolumn{4}{c}{$10^{-4}$ pc$^{-3}$}  
}
\startdata
11.75 &  11.6 $\pm$ 3.1  & 2.84$\pm$0.78  &    2.23$\pm$0.67  & 2.4 \\
12.25 & 8.3 $\pm$ 2.6  &  7.77$\pm$2.40  &     6.60$\pm$2.81  &  8.3 \\
12.75 & 5.0 $\pm$ 2.0  &  4.87$\pm$2.22 &      7.83$\pm$1.28  &  5.0 \\
13.25 & 5.8 $\pm$ 2.2  &  3.77$\pm$2.79 &      7.84$\pm$1.84  &  5.8 \\
13.75 & 5.0 $\pm$ 2.0  &  6.67$\pm$0.44   &    6.62$\pm$0.39  &  5.0 \\
14.25 & $>$6.6 $\pm$ 2.3  & 5.59$\pm$0.73 &  5.66$\pm$0.96 &  6.6 \\
14.75 & $>$3.3 $\pm$ 1.7  & 2.82$\pm$2.51 &  3.47$\pm$1.98 &  3.3 \\
\enddata
\tablecomments{CRUZ7 stands for the LLF from Cruz et al. 2007.
 CRUZ7 is the input for the first $\chi^2$ fitting to generate
 ITERA.
ITERA is the output LLF of CRUZ7 fitting and input LLF
for the confirmation run to check the stability of the solution. 
FINAL is the final LLF after ITERA fitting. HYBRD is the result
of the two parameter fit with the scale height and the $M_J = 11.75$ magnitude bin,
while other magnitude bins are fixed to those of MCRUZ.
 }
\end{deluxetable*}

\subsection{Varying LLF with the Monte Carlo method}

Since the one parameter fit with the scale height alone appears poor,
next we allow each of the seven magnitude bins in the LLF to vary along with
the scale height and this reduces the DOF to 37 ($=45-1-7$).
We understand that some readers may object to increase the parameter space
and we respond to this potential criticism in section \ref{hybrid}.

There are a couple of reasons that we need to investigate the LLF.
First,  the number density error in the currently
known LLF \citep{2007AJ....133..439C} is large 
due to the small number statistics. 
The LLF by Cruz et al. was determined from a total of 55 objects
for $M_J = 11.75 \sim 14.75$ or between M9 and L8 in terms of spectral type
and significant error ($30\sim40$\%) in the number density exists in each magnitude bin.
Second, the number density of stars  in the Galactic thin disk is not uniform (e.g. The Sun is 
in the Orion arm.) and there may be potential idiosyncrasy of the immediate solar 
neighborhood which can only be revealed by analyzing a large number of sufficiently
distant sources detected by a wide and deep survey.
One weakness inherent in a deep survey like ours or HST-based surveys is that
distance estimates are not based on trigonometric parallaxes. 
However even in the 20 pc sample of Cruz et al., only 30\% of the sources have
trigonometric parallaxes, and the distance estimates of the rest of sources
are based on spectrophotometry.
The success or failure of the fitting with varying LLF can only be judged 
objectively by applying the $\chi^2$ distribution for $N=37$ degrees of freedom
or by the figure of merit $Q$, compared to that for the one parameter fit
with the scale height alone.

\begin{figure}
\includegraphics[width=12cm,angle=-90]{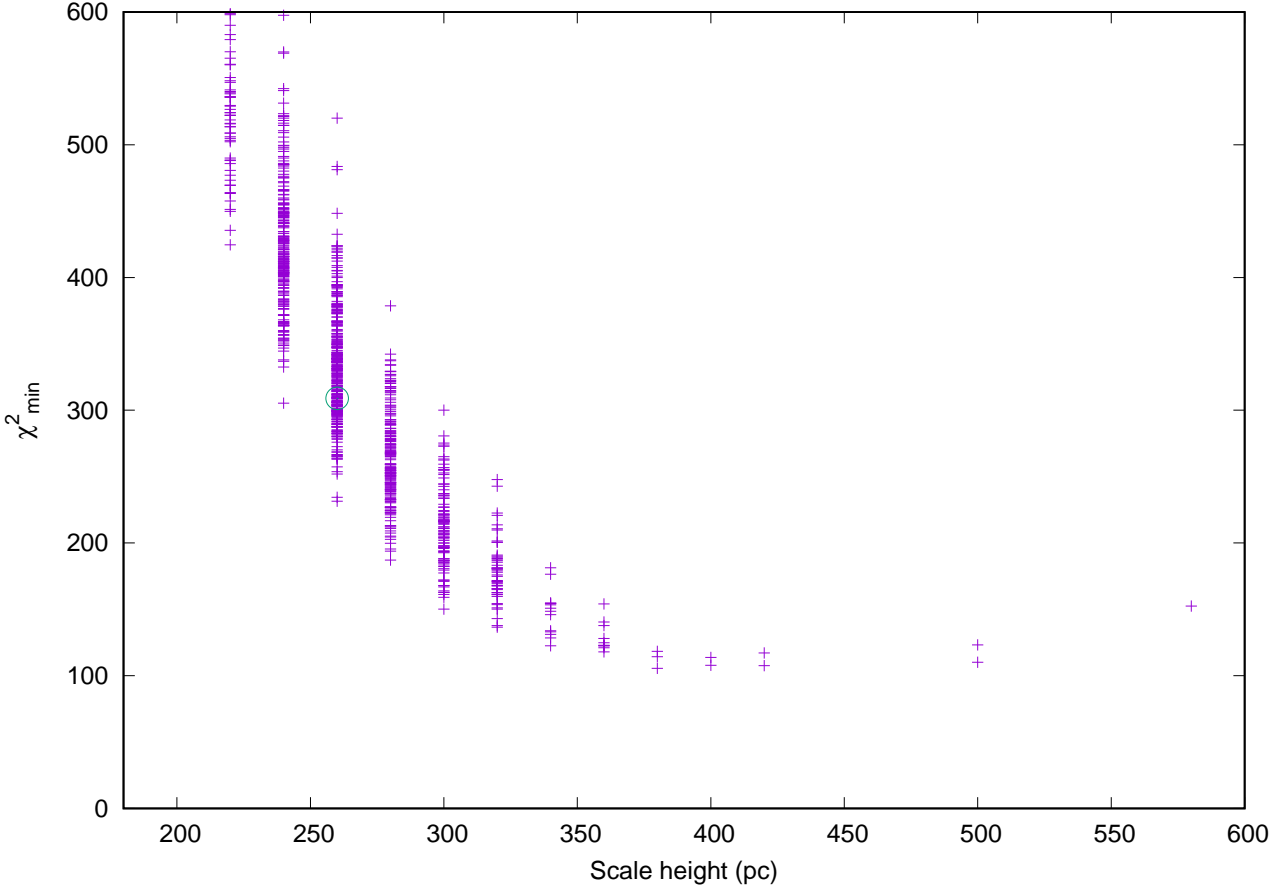}
\caption{Minima of $\chi^2$ and scale heights 
that give the minima for 1000 realizations of LLFs.
The number density for each magnitude bin of each random LLF  is generated 
as a Gaussian random variable with the mean and
standard deviation of the Cruz et al. LLF (CRUZ7).  The mean LLF of Cruz et al. corresponds
to the open circle at ($h$,$\chi^2_{min}$) = (260 pc, 309). The smallest $\chi^2$ 
is located at ($h$,$\chi^2_{min}$) = (380 pc, 106).
The figure of merit $Q=3.0\times10^{-15}$, which is significantly better than the
mean Cruz model.
\label{fig:newchisqF}}
\end{figure}

However, it is very difficult to determine the LLF and $h$ simultaneously
without any guiding principle on the LLF,
since the parameter space is now eight dimensional.
Here we treat the number density for each magnitude bin as
a Gaussian random variable with the mean and standard deviation given
by Cruz et al.. 
To distinguish this LLF with errors from MCRUZ, we call this CRUZ7.
CRUZ7 is given in  
Table \ref{tab:LF}.
Using CRUZ7 as an input,
we generate 1000 LLFs. For each realization
of  LLF, $\chi^2$ is calculated for $h$ between 200 and 580 pc with a 20 pc step
(20,000 models in all)
and $h$ that minimizes $\chi^2$ is searched for. 
In Figure \ref{fig:newchisqF}, the combinations
of the optimal $h$ and $\chi^2_{min}$ are plotted for 1000 realizations of LLFs.
%\textcolor{red}{
%This number of realizations of LLFs, 1000, 
%is to probe the behavior of the optimal $h$ and $\chi^2_{min}$.
%}

%% The "ht!" tells LaTeX to put the figure "here" first, at the "top" next
%% and to override the normal way of calculating a float position
%\begin{figure}[ht!]
%\includegraphics[width=10cm,angle=-90]{Figure5V13.eps}

\begin{figure}
\begin{center}
\includegraphics[width=12cm,angle=-90]{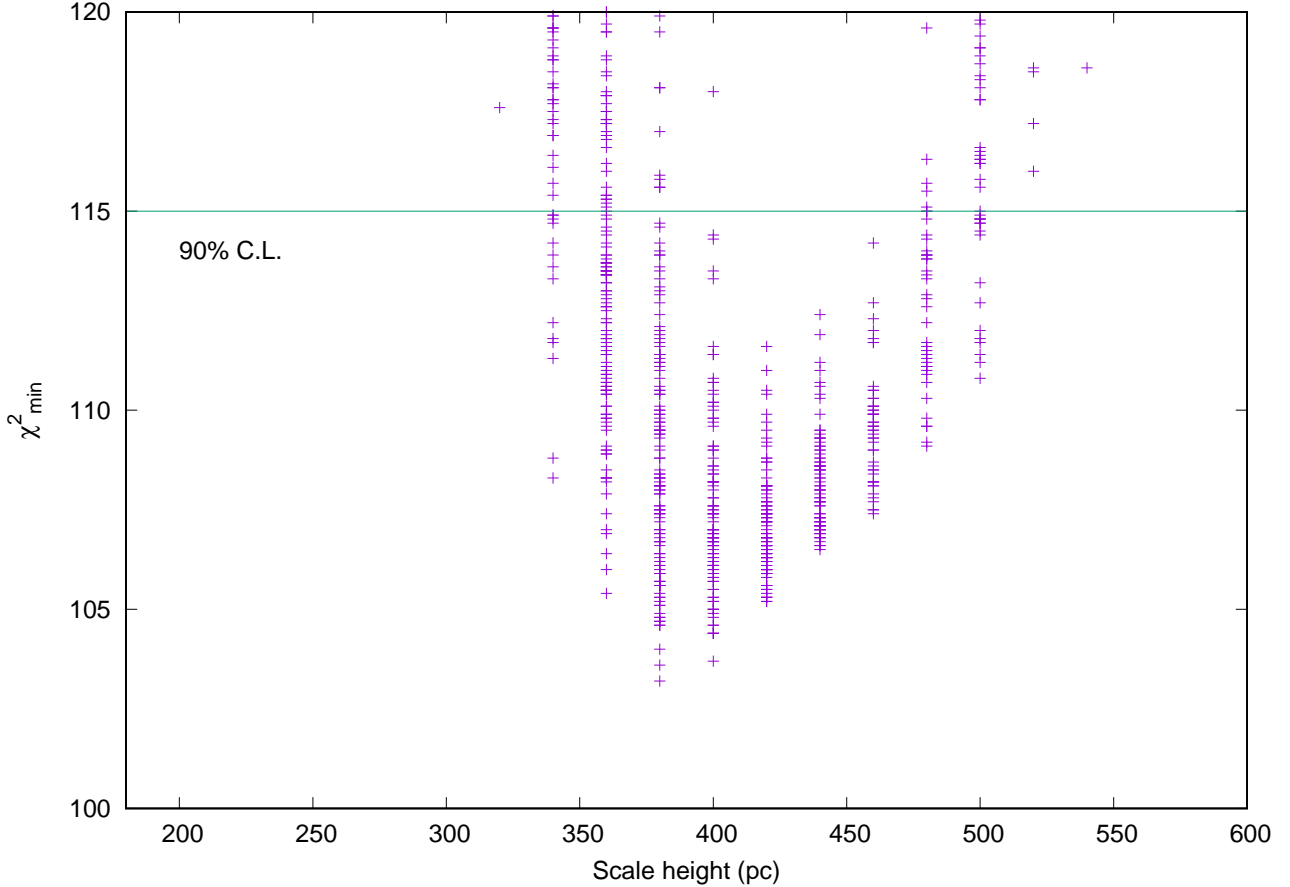}
\end{center}
\caption{ Minima of $\chi^2$ and scale heights for the ITERA model
as input. The global minimum is found at
($h$, $\chi^2_{min}$) = (380 pc, 103), which is only slightly better than
the input model result. This confirms the stability of the solution.
$\Delta \chi^2 = \chi^2_{min} - \chi^2_{min}(global)$ behaves as the 
$\chi^2$-distribution with DOF=8 (the number of free parameters).
The 90\% confidence level is shown as the green horizontal line,
which indicates that the 90\% confidence interval is between 320 
and 520 pc.} 
\label{fig:nitera1D}
\end{figure}

As expected, the number density of data points is largest around the data point
for the MCRUZ model  $(h,\chi^2_{min}) = (260,309)$.
However, there is an interesting trend that $\chi^2_{min}$s are smaller for
larger $h$ and the smallest $\chi^2_{min}$ is  located at 
$(h,\chi^2_{min}) = $(380pc,106) with $Q(37|106) =  3.0\times10^{-15}$.
The $\chi^2_{min}$ appears to increase
beyond $h>400$ pc, but data points are so sparse and nothing definite can be said
about where the true global minimum is.
To investigate the reality of this apparent minimum, we form a LLF from
the LLFs with the three lowest $\chi^2_{min}$ near $h=380$ pc, which
we call ITERA (Table \ref{tab:LF}), meaning iteration.
We use the LLF, ITERA as the seed LLF with the mean
and standard deviation for the $\chi^2$ fitting
and study the stability of the global minimum of $\chi^2_{min}$ found around
$h = 380$ pc.
We generated another 1000 realizations of LLFs and
calculated $\chi^2$ for $h$ from 200 to 580 pc with a 20 pc step and searched for the
optimal $h$ and
$\chi^2_{min}$. The resultant combinations of $h$ and $\chi^2_{min}$ are plotted
in Figure \ref{fig:nitera1D}.  
There is a global minimum of $\chi^2_{min}(global)$ 
of 103 at $h \sim 380$ pc, which gives the figure of merit,  $Q(37|103) = 2.2\times 10^{-14}$,
which is slightly better than the result for ITERA.  This iteration confirms that the global
minimum found at $h=380$ is stable and the search for the solution converged.
The final LLF, FINAL, was formed by averaging LLFs corresponding to the three lowest
$\chi^2$s, and is given in Table \ref{tab:LF}.

\begin{figure}
\begin{center}
\includegraphics[width=12cm,angle=-90]{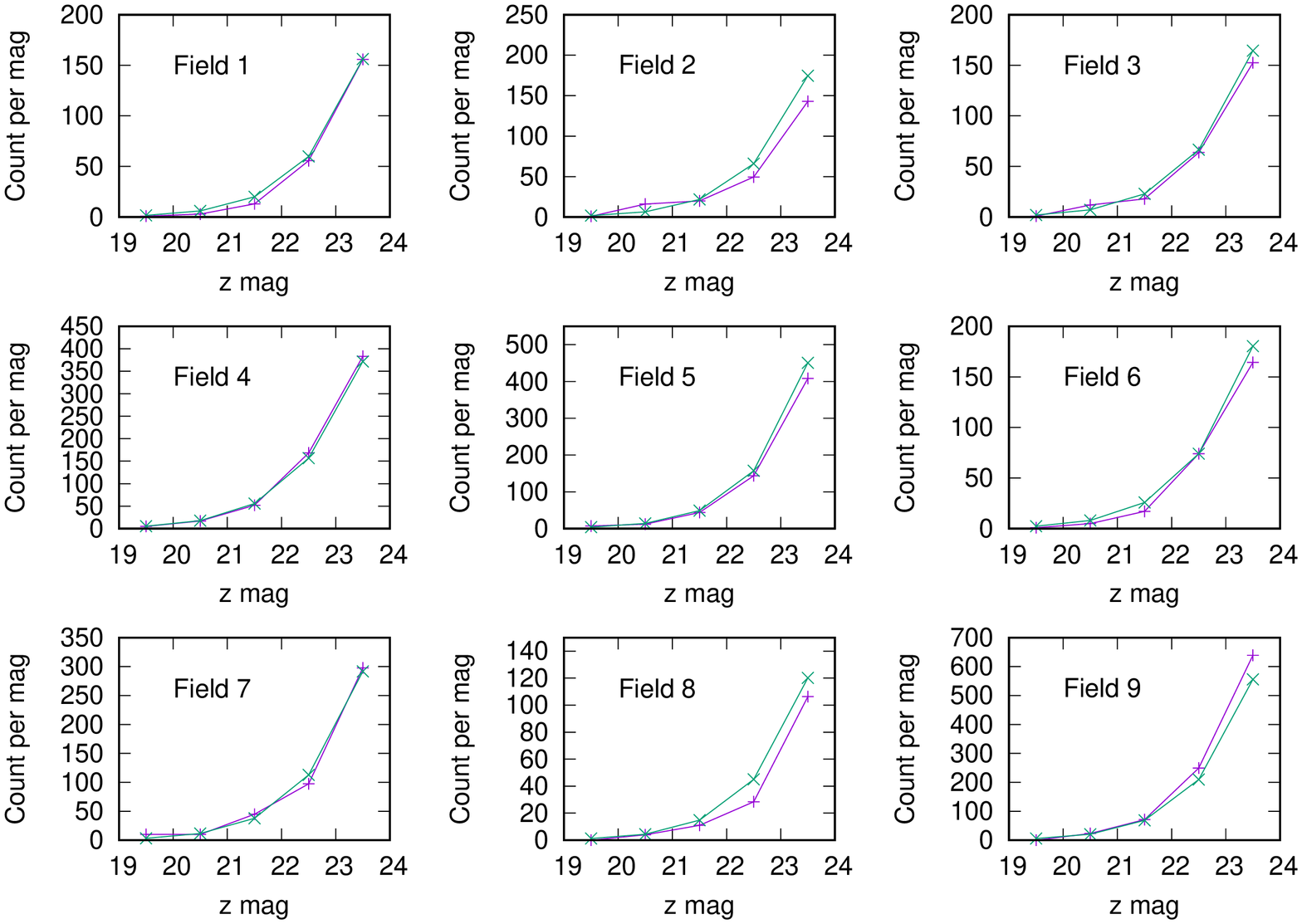}
\end{center}
\caption{Comparison of observed counts (violet) and predictions by the FINAL 
model (green).
The optimal  $h = 380$ pc and $\chi^2 = 103$ for DOF = 37. The figure of merit
$Q = 2.2 \times 10^{-14}$, which is significantly larger than that for the  MCRUZ model.
} 
\label{fig:minLF}
\end{figure}

This global minimum is rather broad as we can see from  the following discussion of statistics.
$\Delta \chi^2 = \chi^2_{min} - \chi^2_{min}(global)$ follows the $\chi^2$ distribution of
eight degrees of freedom (DOF= the number of free parameters to vary) 
\citep{1976ApJ...208..177L}. The $\Delta \chi^2$ for a 90\% confidence 
level is 12. The horizontal line drawn in Figure \ref{fig:nitera1D}, 
shows this level and the data points below this line are within the 90\% confidence
interval.
From the behavior of the lower envelope of data points in Figure \ref{fig:nitera1D},
we can read that the 90\%  confidence interval is between 320 and 520 pc.
This also implies that the MCRUZ model
is ruled out with a very high confidence.
This horizontal line for the 90\% confidence level should be regarded as
an upper bound, because of the following reason. We have assumed
that the densities in individual magnitude bins are independent Gaussian
random variables. However we know that there is a significant error in
the absolute magnitude in each magnitude bin and neighboring bins
are correlated. So effective degrees of freedom of free parameters  should be less than
eight although there is no clue at present to calculate this effective DOF.
The lower DOF implies that the confidence level in Figure \ref{fig:nitera1D} should
be lower and the confidence interval should be narrower. 
The model predictions by the mean number densities of
FINAL and the observed L dwarf count are compared in Figure \ref{fig:minLF}.
We consider that the fit of the observed count with the model predictions is
satisfactory.

\subsection{Comparison of three models}

\begin{figure}
\begin{center}
%\plotone{cmp480m360.eps}
\includegraphics[width=12cm,angle=-90]{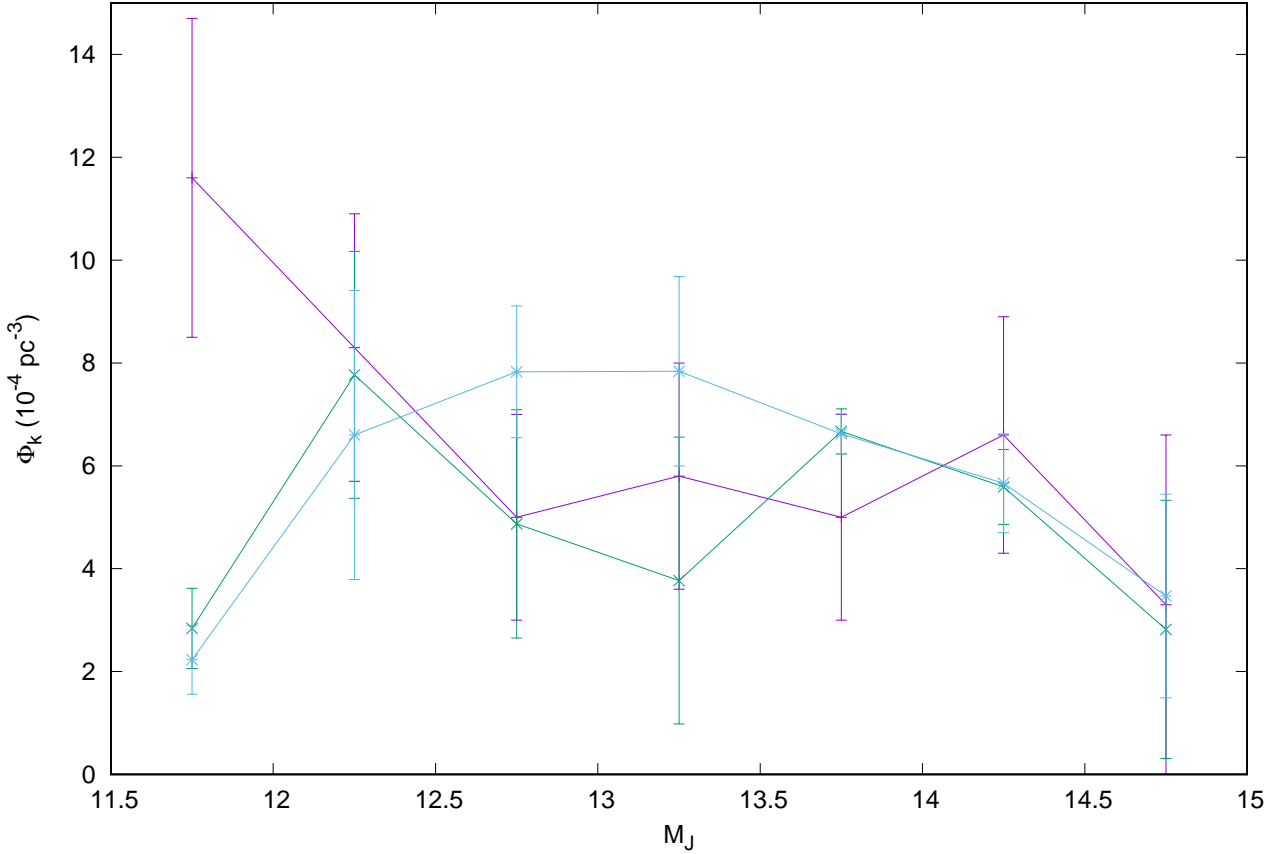}
\end{center}
\caption{Comparison of LLFs,  CRUZ7 (violet), ITERA (green) and FINAL (blue).
The brightest magnitude bin in CRUZ7 is much greater than
those of ITERA and FINAL.  
Apart from this brightest magnitude bin, three luminosity
functions are not significantly different. Error bars are for 1-sigma errors. 
} 
\label{fig:crzsnm}
\end{figure}

Here the three LLFs,  MCRUZ,  ITERA and FINAL are compared.
These LLFs are numerically given in Table \ref{tab:LF}, and plotted in Figure \ref{fig:crzsnm}.

The most striking feature in Figure \ref{fig:crzsnm} is that 
the brightest  magnitude bin in MCRUZ 
 ($M_J = 11.75$) is greater than those of ITERA and FINAL by a
factor of four. 
From the comparison between the observed count and prediction by the MCRUZ model
in Figure \ref{meanLF}, we notice that the MCRUZ model predicts excessive counts
compared to observations in many fields. 
Those bright sources seen in MCRUZ do not exist in the observed count.
The small scale height $h=260$ pc
derived for this model appears to compensate for this excess at large distances from the Sun.
One reason is conceivable for this behavior.
The brightest  bin corresponds to the bluest sources especially in $z-y$ in our sample,
while  the sample of Cruz et al. was selected by spectral type. 
So it is conceivable that
the source counts near the M/L border ($M_J = 11.75$)  may to some extent disagree
due to the difference in the source selection methods.

The second striking feature is that apart from the brightest magnitude bins,
the three models agree reasonably well within the uncertainties
of number densities  of individual
magnitude bins. This is rather surprising considering that the ITERA and FINAL models
are results of the Monte Carlo method.  Except for the brightest magnitude bins
of ITERA and FINAL, MCRUZ was effectively unchanged. 
This indicates that the potential idiosyncrasy of the LLF of the Cruz et al. 
in the immediate solar neighborhood
is small, even if it existed, except for the mystery of the brightest magnitude bin.

\subsection{Hybrid model \label{hybrid}}

From the behavior of LLFs in the previous subsection, we can conjecture that
the brightest magnitude bin is solely responsible for the major behavior of $\chi^2$
and the choice of LLF concerning the other magnitude bins is rather irrelevant.
To investigate this possibility, we have fixed the magnitude bins other than
the brightest one to those of MCRUZ, and performed $\chi^2$ fitting
using the scale height and the density of the brightest magnitude bin 
as the only two free parameters.
Since the parameter space is rather small, we need not to use the Monte Carlo technique.
For the density of the brightest magnitude bin, we calculated 150 data points
between  $1.0 \times 10^{-5}$ and $1.5 \times 10^{-3}$ (pc$^{-3}$) in $1.0\times 10^{-5}$
step and  generated 150 LLFs. 
$\chi^2$s were calculated for $h$ between 200 and 580 pc in a 20 pc step (3000 models),
and $h$, which minimizes the $\chi^2$, was searched for for each 
LLF.  

\begin{figure}
\begin{center}
\includegraphics[width=12cm,angle=-90]{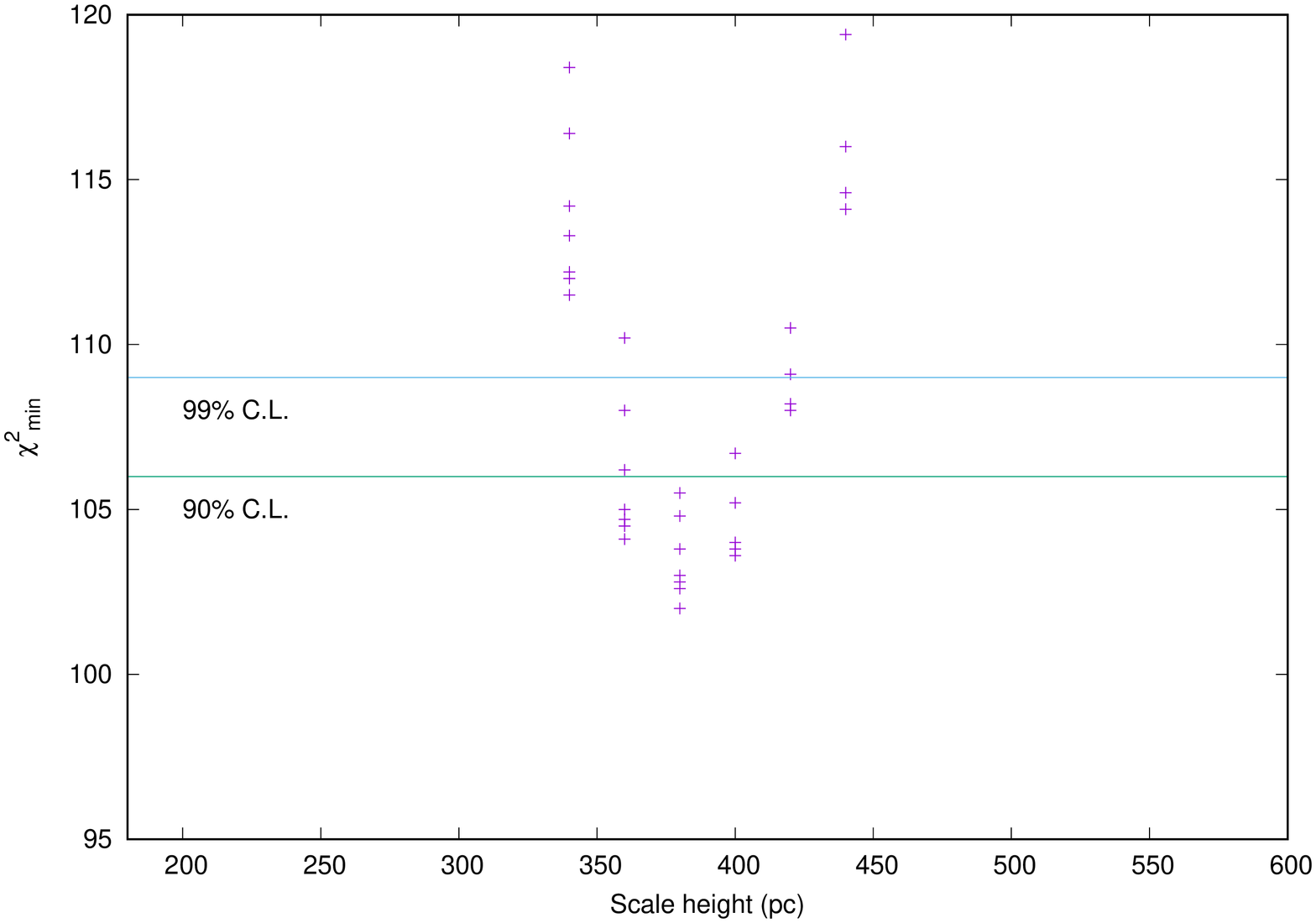}
\end{center}
\caption{ The global minimum (the HYBRD model) is found at
($h$, $\chi^2_{min}$) = (380 pc, 102), which is 
exactly the same for FINAL.
$\Delta \chi^2 = \chi^2_{min} - \chi^2_{min}(global)$ behaves as the 
$\chi^2$-distribution with DOF=2.
The 90\% and 99\% confidence levels are shown as the green horizontal lines.
} 
\label{fig:hybrid}
\end{figure}

The result confirms our anticipation. The global minimum of $\chi^2$, 
was found at $(h,\chi^2_{min}) = $(380 pc, 102), which is exactly the same
as for the global minimum found for the Monte Carlo technique. 
We call this model, the HYBRD model (hybrid model), 
since the LLF for this model is partly MCRUZ and partly a random
LLF.  The number density for the brightest magnitude bin of HYBRD is
$2.4 \times 10^{-4}$ pc$^{-3}$.
The figure of merit is $Q(43|102) = 1.4 \times 10^{-11}$, which is greater than
that for the global minimum of the random LLF.
The behavior of $\chi^2_{min}$ is plotted in Figure \ref{fig:hybrid}.
Although the location of the minimum is the same, the width of
the minimum is much narrower than the case of the fully random LLF.
This is because we did not vary the lower six magnitude bins.
$\Delta \chi^2 = \chi^2_{min} - \chi^2_{min}(global)$ follows 
$\chi^2$ probability distribution of only two degrees of freedom.
A 90\% confidence interval is between 340 and 420 pc, while
a 99\% confidence interval is between 340 and 440 pc.
This indicates that  the vertical scale height for the global minimum, 380 pc,
is rather secure.

%The number densities for the faintest two magnitude bins ($M_J = 14.25$ and $14.75$) of the MCRUZ model %are given as lower limits. We should perhaps comment on why the number densities for these magnitude bins %even after comparing with observations do not change.
%Cruz et al. considered that the number densities are lower limits not because   data are incomplete
%as the number densities of  L dwarfs, but because L/T transition objects which have  $M_J$ in these
%magnitude bins due to $J$ brightening are spectroscopically excluded.
%In our sample, L/T transition objects are also excluded by $i-z$ and $z-y$ colors and 
%the comparison of the MCRUZ model and observations is self-consistent.

\subsection{Effect of binaries}

Here we try to evaluate the upper bound of the influence of binaries on our analysis.
By NICMOS/HST imaging, \citet{2008AJ....135..580R} obtained a binary fraction of L dwarfs
within 20 pc to be $12.5^{+5.3}_{-3.0}$\% after correcting for  Malmquist bias.
Taking into account the fact that there may be some binaries spatially unresolved even
by HST, we here assume a binary fraction of 20\% as an upper bound.
Equal flux binaries have the maximum effect on the source count and we assume
that all binaries are equal flux binaries.
From the original source catalog, we generate a new catalog by
assuming that each source is an equal flux binary with a 20\% probability.
We generate multiple new catalogs and average them to make a 
catalog containing binaries. 
The usual   $\chi^2$ fitting is performed for the catalog containing binaries.
The global minimum of $\chi^2_{min} = 105$ is found at $h \sim 380$ pc.
The LLF for this global minimum coincides with FINAL within errors.
It turned out that the effect of binaries is smaller than that of
the completeness correction mentioned above.

\begin{table}
\begin{center}
\begin{tabular}{ccccc}
\tableline\tableline  
  & MCRUZ & ITERA & FINAL  & HYBRD \\
\tableline
   $h$(pc)  &   260  & 380 & 380 & 380  \\  
DOF   &   44   &   37  &  37  & 43 \\
$\chi^2_{min}$ & 309 & 106 & 103 & 102  \\
$Q$       &  $1.4\times 10^{-80}$  &   $3.0\times10^{-15}$ & $2.2\times10^{-14}$ 
& $1.4\times10^{-11}$\\
\tableline
\end{tabular}
\end{center}
\caption{Results of fitting. The best-fit models are compared.
MCRUZ stands for the model for the mean LLF of Cruz et al.(2007).
ITERA is the result after varying the Cruz LLF (CRUZ7).
FINAL is the result of the confirmation run using the ITERA model as input.
HYBRD is the LLF with varying the brightest magnitude bin and static other bins
fixed to those of MCRUZ.
$h$ is the optimal scale height.  DOF is the degrees of freedom of fitting. $Q$ is the figure of 
merit for fitting to compare the goodness of fit for different degrees of freedom. Larger the $Q$,
better the fit.}
\end{table}

\section{Discussion}

\subsection{Comparison with previous works on brown dwarfs}

As we have shown in Figure \ref{fig:comparison} (Comparison of ground-based surveys),
previous ground-based wide-field surveys have not been able to detect L dwarfs
at 300 pc. Therefore there has not been any attempt to estimate
the vertical scale height based on the ground-based data.
HST observations are deep enough to reach brown dwarfs well beyond 300 pc,
but suffer from the small field of view and therefore from small number statistics.
Estimates of the brown dwarf scale height are summarized in Table \ref{tab:BDhz}.

Based on 18 M and later dwarfs, \citet{2005ApJ...622..319P}
determined the vertical scale height of M and L dwarfs to be 400 $\pm$ 100 pc.
For the purpose of comparing observations with Galaxy models
they used model luminosity functions derived from Monte Carlo mass function
simulations developed by \citet{2004ApJS..155..191B} . 

From 28 L and T dwarfs selected by $i^\prime-z^\prime$ color and morphology,
\citet{2005ApJ...631L.159R} estimated the scale height of L and T dwarfs to be
350 $\pm$ 50 pc. 
In their analysis, L and T dwarfs were treated as a single population and 
a local number density for L and T dwarfs derived by \citet{2002ApJ...567..304C}
was used.

By infrared imaging, \citet{2011ApJ...739...83R}  discovered
17 ultracool dwarfs later than $\sim$M8 and estimated the vertical
scale height to be 290 $\pm$ 25 (random) $\pm$ 31 (systematic).
It should be noted that they used the LLFs of \citet{2007AJ....133..439C}
for M8-L8 dwarfs and of \citet{2010A&A...522A.112R} for T dwarfs.  They took the uncertainties
in the LLFs into account as we have done in the previous section and
computed the systematic error. 
It is surprising that errors in the scale height are so small despite the small sample size of 17 sources.
We suspect that the treatment of statistics in estimating the error due to uncertainties
in the LLFs may not be appropriate. When individual magnitude bins
in the LLFs are varied along with the scale height,
there are more than ten DOF in $\Delta \chi^2 = \chi^2 - \chi^2_{min}$ distribution.
This large dimension in the parameter space may result in a greater statistical error.
Although there is some question in the error estimate, the work by  \citet{2011ApJ...739...83R}
should be compared with ours from the point of view of the level of sophistication
in analysis. Their result in $h = 290$ pc is significantly smaller than
 our scale height estimate.

\begin{deluxetable*}{cccc}[b!]
\tablecaption{Scale height estimates of brown dwarfs\label{tab:BDhz}}
\tablecolumns{4}
%\tablenum{2}
\tablewidth{0pt}
\tablehead{
\colhead{Authors} &
\colhead{Facility}  & \colhead{Scale height}  \\
 & & \colhead{pc}  & \colhead{Comment}
}
\startdata
Pirzkal et al. (2005) & HST & 400 $\pm$ 100 & Simulated LF  \\
Ryan et al. (2005) & HST & 350 $\pm$ 50    &  L \& T dwarfs treated as a single population\\
Ryan et al. (2011) & HST & 290 $\pm$ 25(random) $\pm$ 31 (systematic)  & \\
This work(FINAL)           & HSC & 380$^{+140}_{-60}$ & 90\% confidence interval \\
This work(HYBRD)           & HSC & 380$\pm$ 40            & 90\% confidence interval \\ 
\enddata
%\tablecomments{
%}
\end{deluxetable*}

\begin{deluxetable*}{cccc}[b!]
\tablecaption{Scale height estimates of low mass stars\label{tab:Mdwarfhz}}
\tablecolumns{4}
%\tablenum{2}
\tablewidth{0pt}
\tablehead{
\colhead{Authors} &
\colhead{Facility}  & \colhead{Scale height} & Comment  \\
 & & \colhead{pc} 
}
\startdata
Chen et al. (2001) & SDSS & 330 $\pm$ 3  & Fixed LF \\
Juri\'c et al. (2008) & SDSS & 300 $\pm$ 60 & \\
Bochanski et al. (2010) & SDSS & 300 $\pm$ 15 & \\
Pirzkal et al. (2009) & HST & 370$^{+60}_{-65}$ &  M4-M9 dwarfs \\
Pirzkal et al. (2009) & HST  & 300 $\pm$ 70 &    M0-M9 dwarfs \\
van Vledder et al. (2016) & HST  & 290$^{+20}_{-19}$ &  \\
\enddata
%\tablecomments{
%}
\end{deluxetable*}

\subsection{Comparison with scale height estimates of low mass stars}

To compare with the brown dwarf scale height, scale height estimates of low mass stars
are summarized in Table \ref{tab:Mdwarfhz}.
The most recent determinations of the vertical scale height of low mass stars
are based on the SDSS data \citep{2001ApJ...553..184C,2008ApJ...673..864J,2010AJ....139.2679B}.
Chen et al. used the LLF of \citet{1983nssl.conf..163W} and derived a thin-disk scale height
of $h = 330 \pm 3$ pc for late-type dwarfs. Their error is so small because
they adopted a fixed LLF.
Juri\'{c} et al. used the photometric parallax method to estimate the distances
to stars and mapped their three-dimensional number density distribution
in the Galaxy. From Mote Carlo simulations, they estimated the errors of $\sim$20\%
in the disk scales and $\sim$10\% in the density normalization. They derived
a M dwarf scale height of $h = 300 \pm 60$ pc.  They consider that
the largest contributions to error come from the uncertainty in calibration
of the photometric parallax relation and poorly constrained binary fraction.
Bochanski et al. simultaneously measured Galactic structure and the stellar LF 
for stars with masses between 0.1 and 0.8M$_\odot$.
They derived a thin disk scale height of $h = 300 \pm 15$ pc.
Despite using somewhat different approaches,
Juri\'{c} et al. and Bochanski et al.
derived the same mean value of $h = 300$ pc, which appears to represent
the results from SDSS for low mass stars.

The estimation of the vertical scale height of M dwarfs has also been attempted
by the use of the HST data \citep{2009ApJ...695.1591P,2016MNRAS.458..425V}.
Based on the sources selected from
 the ACS grism spectra, Pirzkal et al. estimated the vertical scale height of
 $h = 370^{+60}_{-65}$ pc for M4-M9 dwarfs. They also derived the scale height for
M0-M9 dwarfs  to be $h  = 300 \pm 70$ pc. In their analysis, they adopted
the $z$-band LF of  \citet{2010AJ....139.2679B} and the $J$-band LF of
\citet{2007AJ....133..439C}.
van Vledder et al. selected 274 identified M dwarfs in the WFPC3 pure
parallel fields from the BoRG survey for high-redshift galaxies.
Although they use the terminology ``M-type brown dwarfs", what they mean by that appears
to be ordinary M dwarfs.
They derived the disk scale height of $h = 290^{+20}_{-19}$ pc.
As the scale height of M dwarfs (M0-M9), the results from the HST data
also have a mean of the scale height around $h \sim 300$ pc, which is in
accord with the results from the SDSS data.

The scale height of L dwarfs we derived from the HSC data 
(especially the HYBRD model)
appears 
higher than  the mean value for M dwarfs.

\subsection{Comparison with kinematics of nearby dwarfs}

One way to study the structure of the Galaxy is to investigate the kinematics
of low mass  stars and substellar objects in the immediate solar neighborhood.
A study of tangential velocities for a large sample of late-type M, L, and T dwarfs
has found that the kinematics of the sample show no significant differences
between late-type M, L, and T dwarfs \citep{2009AJ....137....1F}.
A recent study of $UVW$ velocities of late M dwarfs and L dwarfs shows interesting
and mysterious results \citep{2015ApJS..220...18B}. While the simulations based on evolutionary models
predict younger ages for L dwarfs than for M dwarfs, observations  indicate
older ages for L dwarfs (6.5 $\pm$ 0.4 Gyr) than for M dwarfs (4.0 $\pm$ 0.2 Gyr).
The scale height we derived is qualitatively in accord with the results by Burgasser et al.,
since the scale height for L dwarfs (380 pc) appears to be larger than that for M dwarfs (300 pc).

\section{Concluding remarks}

We have analyzed DR1 of the HSC-SSP data aiming at determining 
the vertical scale height of L dwarfs in the Galactic thin disk, assuming
an exponential disk model.
We first used the mean LLF of \citet{2007AJ....133..439C}, but this LLF
resulted in a poor fit to the data even for an optimal scale height of $h = 260$ pc.
We then allowed the number densities for individual seven magnitude bins to vary
along with the scale height, which is an eight parameter fit. We reached a reasonable fit for a scale
height of $h = 380$ pc.
However, this minimum is broad
and a 90 \% confidence interval  is between 320 and 520 pc.
Then we restricted the free parameters to the scale height and the density of
the brightest magnitude bin in the LLF, which is a two parameter fit. 
We reached an equally good fit to the eight parameter fit with the two parameter fit
and a 90\% confidence interval was between 340 and 420 pc. 
A deeper limiting magnitude is necessary to study the scale height of T dwarfs.
The HSC  ultra-deep survey will reach 1.7 magnitude deeper than the wide survey,
although the survey area  is 3.5 deg$^2$.
We plan to pursue the possibility of utilizing this survey aiming at estimating
the T dwarf scale height.

%% If you wish to include an acknowledgments section in your paper,
%% separate it off from the body of the text using the \acknowledgments
%% command.
\acknowledgments

We thank the anonymous referee for critical reading of the manuscript and
useful suggestions. We acknowledge Drs. Adam Burgasser,   Sandy Leggett
and Michael Cushing for providing us with observed near-infrared spectra.
We are grateful to everyone involved in the hardware development, observations,
and data reduction for the HSC-SSP survey. This work was supported by
Grants-in-Aid for Scientific Research from the Japan Society for the Promotion
of Science (JSPS) and by the National Astronomical Observatory of Japan (NAOJ) .
 Y. Matsuoka was supported by JSPS KAKENHI Grant No.
JP17H04830 and the Mitsubishi Foundation Grant No. 30140.

The Hyper Suprime Cam (HSC) collaboration includes the astronomical communities
of Japan, Taiwan, and Princeton University. The HSC instrumentation and software
were developed by the National Astronomical Observatory of Japan (NAOJ),
the Kavli Institute for the Physics and Mathematics of the Universe (Kavli IPMU),
the University of Tokyo, the High Energy Accelerator Research Organization (KEK),
the Academia Sinica Institute for Astronomy and Astrophysics in Taiwan (ASIAA),
and Princeton University.
Funding was contributed by the FIRST program from Japanese Cabinet Office,
the Ministry of Education, Culture, Sports, Science and Technology (MEXT),
the Japan Society for the Promotion of Science (JSPS), Japan Science and
Technology Agency (JST), the Toray Science Foundation, NAOJ, Kavli IPMU,
KEK, ASIAA, and Princeton University.

%This paper makes use of software developed for the Large Synoptic Survey Telescope.
%We thank the LSST Project for making their code available as free software at
%http://dm.lsst.org

%\noindent
%\textcolor{red}{Pan-STARRS is acknowledged in Sorahana san's manuscript.
%Is this necessary? If so, please add here.} 

%% To help institutions obtain information on the effectiveness of their 
%% telescopes the AAS Journals has created a group of keywords for telescope 
%% facilities.
%
%% Following the acknowledgments section, use the following syntax and the
%% \facility{} or \facilities{} macros to list the keywords of facilities used 
%% in the research for the paper.  Each keyword is check against the master 
%% list during copy editing.  Individual instruments can be provided in 
%% parentheses, after the keyword, but they are not verified.

%% Appendix material should be preceded with a single \appendix command.
%% There should be a \section command for each appendix. Mark appendix
%% subsections with the same markup you use in the main body of the paper.

%% Each Appendix (indicated with \section) will be lettered A, B, C, etc.
%% The equation counter will reset when it encounters the \appendix
%% command and will number appendix equations (A1), (A2), etc. The
%% Figure and Table counter will not reset.

\appendix

\section{Explicit treatment of the Malmquist bias}

Here we treat the case of one particular stellar population,
whose volume number density of the solar neighborhood is $n_0$, whose mean absolute
magnitude is $M_0$ and its dispersion is $\sigma$.
We follow the notation of the original work by \citet{1922MeLuF.100....1M}.

If we denote by $a(m)$ the frequency function of the apparent magnitude
$m$, then $a(m)dm$ is the number of stars having an apparent magnitude
$m \pm dm/2$ and

\begin{equation}
\int_{m-1/2}^{m+1/2} dm a(m) = N(m),
\end{equation}

is the differential number count at the apparent magnitude $m$ in units of 
the number per mag. If the relative frequency function of the absolute magnitude,
$\varphi(M)$ has a Gaussian form as assumed by Malmquist,

\begin{equation}
\varphi(M) = \frac{1}{\sqrt{2\pi \sigma^2}} \exp\left(- \frac{(M-M_0)^2}{2 \sigma^2}\right),
\end{equation}

where $\int dM \varphi(M) = 1$, 
and the volume number density at distance $r$ from us is $D(r)$, $a(m)$ is given by

\begin{equation} 
a(m) = \omega \int_0^{+\infty} dr r^2 D(r) \varphi(M),
\end{equation}

and the differential number count is

\begin{equation}
N(m) = \omega \int_{m-1/2}^{m+1/2} dm \int_{0}^{+\infty} dr r^2 D(r) \varphi(M),
\end{equation}

where $\omega$ is the solid angle of the area of concern.
In our exponential disk model,

\begin{equation}
D(r) \varphi(M) = n_0 \exp\left(\frac{R_0-R}{H} - \frac{|Z|}{h}\right) \frac{1}{\sqrt{2 \pi \sigma^2}}
       \exp\left(-  \frac{(M-M_0)^2}{2 \sigma^2}\right),
\end{equation}

and

\begin{eqnarray}
R(r,b,l) & = & \sqrt{R_0^2 + (r \cos b)^2 - 2R_0 r \cos b \cos l}, \\
Z(r,b) & = & Z_\odot + r \sin(b - \tan^{-1}(Z_\odot / R_0)), \\
M(m,r) & = & m - 5 \log r +5,
\end{eqnarray}
               
where $Z_\odot$ is the height of the Sun from the Galactic plane.
The effect of the Malmquist bias is explicitly taken into account by the numerical integration in 
the apparent magnitude $m$.

\end{document}